\documentclass[]{aastex631}

\usepackage[figuresright]{rotating}
\shorttitle{AASTeX v6.3.1 Sample article}
\shortauthors{Ding et al.}
\graphicspath{{./}{figures/}}

\begin{document}

\title{Fundamental Parameters for Totally Eclipsing Contact Binaries Observed by TESS}
\correspondingauthor{KaiFan Ji}
\email{jkf@ynao.ac.cn}

\author[0000-0002-2427-161X]{Xu Ding}
\affiliation{Yunnan Observatories, Chinese Academy of Sciences (CAS), P.O. Box 110, 650216 Kunming, P. R. China}
\affiliation{Key Laboratory of the Structure and Evolution of Celestial Objects, Chinese Academy of Sciences, P. O. Box 110, 650216 Kunming, P. R. China}
\affiliation{Center for Astronomical Mega-Science, Chinese Academy of Sciences, 20A Datun Road, Chaoyang District, Beijing, 100012, P. R. China}

\author{KaiFan Ji}
\affiliation{Yunnan Observatories, Chinese Academy of Sciences (CAS), P.O. Box 110, 650216 Kunming, P. R. China}
\affiliation{Key Laboratory of the Structure and Evolution of Celestial Objects, Chinese Academy of Sciences, P. O. Box 110, 650216 Kunming, P. R. China}
\affiliation{Center for Astronomical Mega-Science, Chinese Academy of Sciences, 20A Datun Road, Chaoyang District, Beijing, 100012, P. R. China}
\affiliation{University of the Chinese Academy of Sciences, Yuquan Road 19\#, Shijingshan Block, 100049 Beijing, P.R. China}

\author{ZhiMing Song}
\affiliation{School of Information, Yunnan University of Finance and Economics, Kunming, China.}
\affiliation{Yunnan Key Laboratory of Service Computing, Kunming, China.}

\author{XueFen Tian}
\affiliation{Communication And Information Engneering College, Yunnan Open University, 650500 Kunming, P. R. China}

\author{JinLiang Wang}
\affiliation{Yunnan Observatories, Chinese Academy of Sciences (CAS), P.O. Box 110, 650216 Kunming, P. R. China}
\affiliation{Key Laboratory of the Structure and Evolution of Celestial Objects, Chinese Academy of Sciences, P. O. Box 110, 650216 Kunming, P. R. China}
\affiliation{Center for Astronomical Mega-Science, Chinese Academy of Sciences, 20A Datun Road, Chaoyang District, Beijing, 100012, P. R. China}
\affiliation{University of the Chinese Academy of Sciences, Yuquan Road 19\#, Shijingshan Block, 100049 Beijing, P.R. China}

\author{ChuanJun Wang}
\affiliation{Yunnan Observatories, Chinese Academy of Sciences (CAS), P.O. Box 110, 650216 Kunming, P. R. China}
\affiliation{Key Laboratory of the Structure and Evolution of Celestial Objects, Chinese Academy of Sciences, P. O. Box 110, 650216 Kunming, P. R. China}
\affiliation{Center for Astronomical Mega-Science, Chinese Academy of Sciences, 20A Datun Road, Chaoyang District, Beijing, 100012, P. R. China}
\affiliation{University of the Chinese Academy of Sciences, Yuquan Road 19\#, Shijingshan Block, 100049 Beijing, P.R. China}

\author{QiYuan Cheng}
\affiliation{Yunnan Observatories, Chinese Academy of Sciences (CAS), P.O. Box 110, 650216 Kunming, P. R. China}
\affiliation{Key Laboratory of the Structure and Evolution of Celestial Objects, Chinese Academy of Sciences, P. O. Box 110, 650216 Kunming, P. R. China}
\affiliation{Center for Astronomical Mega-Science, Chinese Academy of Sciences, 20A Datun Road, Chaoyang District, Beijing, 100012, P. R. China}
\affiliation{University of the Chinese Academy of Sciences, Yuquan Road 19\#, Shijingshan Block, 100049 Beijing, P.R. China}

\author{JianPing Xiong}
\affiliation{Yunnan Observatories, Chinese Academy of Sciences (CAS), P.O. Box 110, 650216 Kunming, P. R. China}
\affiliation{Key Laboratory of the Structure and Evolution of Celestial Objects, Chinese Academy of Sciences, P. O. Box 110, 650216 Kunming, P. R. China}
\affiliation{Center for Astronomical Mega-Science, Chinese Academy of Sciences, 20A Datun Road, Chaoyang District, Beijing, 100012, P. R. China}



\begin{abstract}
Totally eclipsing contact binaries provide a unique opportunity to accurately determine mass ratios through photometric methods alone, eliminating the need for spectroscopic data. Studying low mass ratio (LMR) contact binaries is crucial for advancing our understanding of binary star evolution and the formation of rare optical transients known as red novae. 
We identified 143 totally eclipsing contact binaries from TESS. These high-precision light curves reveal a distinct O'Connell effect, which we interpret by introducing a cool spot on the primary star. Training a neural network model that includes a cool spot parameters can generate a high-precision light curve two orders of magnitude faster than Phoebe. Utilizing the neural network (NN$_{nol3}$) model combined with the Markov Chain Monte Carlo (MCMC) algorithm, we rapidly derived the fundamental parameters of these systems. By leveraging the relationship between orbital period and semi-major axis using Random Sample Consensus (RANSAC) algorithm, we estimated their absolute parameters. Our analysis identified 96 targets with mass ratios below 0.25, all of which were not listed in any previous catalog, thus signifying the discovery of new low mass ratio system candidates. Assuming all 143 binary systems are affected by a third light during parameter estimation, we train a neural network (NN$_{l3}$) model considering the third light. Then we calculate the residuals between the mass ratio $q_{l3}$ (considering the third light) and $q_{nol3}$ (neglecting it). For these residuals, the 25th percentile ($Q_1$) is 0.012, the median ($Q_2$) is 0.026, and the 75th percentile ($Q_3$) is 0.05. 

\end{abstract}



\section{Introduction} 
The two component stars in a contact binary system are completely within their respective Roche lobes \citep{Kopal+et+al+1959,Lucy+et+al+1968a,Lucy+et+al+1968b}, sharing a common envelope and undergoing significant mass exchange \citep{Lucy+et+al+1979}. Despite potential substantial mass discrepancies between the two stars, their temperatures remain relatively similar \citep{Kuiper+et+al+1941}. These parameters, including mass ratio, fill-out factor, and temperature ratio, are crucial for understanding the various evolutionary stages of contact binaries \citep{Yakut+et+al+2005,Yildiz+et+al+2013,Li+et+al+2020,Sun+et+al+2020,Latkovi+et+al+2021}. To accurately determine the fundamental parameters of a contact binary, both photometric light curves and radial velocity data are necessary. Given the necessity of a considerably larger telescope for radial velocity observations compared to photometric studies, only a limited number of contact binaries have their radial velocities measured. \citet{Pribulla+et+al+2003} determined that contact binaries displaying flat-bottom minima in their light curves possess photometric mass ratios nearly identical to their spectroscopic counterparts, indicating that total eclipsing contact binaries can achieve reliable mass ratio determinations without spectroscopic data. The behavior of totally eclipsing W UMa systems, as anticipated by the Lucy model, strongly constrains their geometry and the models fitted to their light curves \citep{Mochnacki+et+al+1972a,Mochnacki+et+al+1972b}. Reliable mass ratio parameters are essential for identifying contact binaries with low mass ratios. Theoretically, it is suggested that contact binaries exhibit a low mass ratio cutoff and are prone to merging into fast-rotating single stars due to the Darwin instability \citep{Rasio+et+al+1995, Li+et+al+2006, Arbutina+et+al+2007, Arbutina+et+al+2009, Jiang+et+al+2010, Wadhwa+et+al+2021}. As of now, the only confirmed merging event involving contact binaries, based on observational data, is that of V1309 Sco \citep{Tylenda+et+al+2011}.

The proliferation of light curve data from surveys such as the Catalina Sky Survey (CSS) \citep{Marsh+et+al+2017}, the Kepler mission \citep{Borucki+et+al+2010, Koch+et+al+2010}, and the Transiting Exoplanet Survey Satellite (TESS) \citep{Ricker+et+al+2015} has facilitated the identification of total eclipsing contact binaries with low mass ratios \citep{Christopoulou+et+al+2022, Lalounta+et+al+2024}. Notably, \citet{Ding+et+al+2024} utilized an autoencoder neural network to sift through TESS survey data, uncovering 1322 candidate contact binaries. This catalog provides highly precise light curves of total eclipsing contact binaries observed by TESS. Traditional methods for deriving contact binary parameters, such as the Wilson–Devinney program \citep{Wilson+et+al+1990,Wilson+et+al+2012,Wilson+et+al+1971,Wilson+et+al+2010,Van+et+al+2007} and the Phoebe program \citep{prisa+et+al+2016}, are time-consuming, often requiring hours to days per target. Consequently, the derivation of parameters for a large number of light curves remains a formidable challenge. Machine learning techniques have shown promise in accelerating parameter acquisition for binaries \citep{prsa+et+al+2008, Ding+et+al+2022, Xiong+et+al+2024}. However, these methods typically rely on symmetric light curves and do not account for the O'Connell effect \citep{Connell+et+al+1951, Milone+et+al+1968}, which refers to asymmetric light curves.

In this study, we introduce a novel approach that establishes a mapping from parameters, including a cool spot parameters, to light curves. We train a neural network (NN$_{nol3}$) model and subsequently employ the Markov Chain Monte Carlo (MCMC) algorithm from the emcee program \citep{Foreman+et+al+2019} to swiftly obtain the posterior distribution of the parameters. This methodology effectively fits the light curves of contact binaries exhibiting the O'Connell effect. Ultimately, we compile a catalog of 143 total eclipsing contact binary parameters, 96 of which represent newly identified low mass ratio contact binary candidates. Given that all 143 binary systems are impacted by a third light source during parameter estimation, a neural network model (NN$_{l3}$) incorporating this third light was trained. The mass ratios $q_{l3}$ (considering the third light) and $q_{nol3}$ (neglecting it) were compared. The structure of this paper is as follows. Section 2 outlines the sample selection process. Section 3 details the establishment of the neural network model. Section 4 presents the derivation of parameters for the total eclipsing contact binaries. Section 5 includes discussions, primarily comparing our findings with other catalogs. The final section concludes the paper.

\section{TESS sample selection} 
\citet{Ding+et+al+2024} employed an autoencoder neural network to search for contact binaries from TESS, successfully identifying 1322 candidate contact binaries. We utilize the Python package lightkurve \footnote{https://docs.lightkurve.org/} to obtain the TESS light curves authored by SPOC for these targets. We employed the Lomb-Scargle periodogram \citep{Lomb+1976,Scargle+1982} along with the bootstrap \citep{Efron+et+al+1979} method to estimate the orbital periods and their associated uncertainties for these targets. The light curves exhibit the O’Connell effect, characterized by unequal brightness levels of the maxima occurring between the eclipses. Therefore, we only folded the light curves for the first three periods. To transform the flux values obtained from TESS into magnitudes, we apply the following formulas.
\begin{equation}
    mag_i = -2.5 \times log10(flux_i)  
\end{equation}
\begin{equation} 
    \Delta{mag_i} = mag_i - \frac{\sum_{i=1}^n mag_i}{n}
\end{equation}

where $flux_i$ is the flux obtained from TESS. $\Delta{mag_i}$ is the normalized value.

We identified 143 total eclipsing contact binaries among these targets whose light curves exhibit flat-bottom minima. The application of Gaussian Process Regression (GPR) \footnote{https://github.com/dfm/george/} for accurately fitting the light curve. During the first and second out-of-eclipse phases, the maximum magnitudes are denoted as $Max.I$ and $Max.II$ respectively. Taking TIC 103656297 as an example, its relevant data are presented in the left panel of Figure 1. The results for the 143 targets show that the value of $\Delta M_{ag}$, which is calculated as $\Delta M_{ag} = Max.I - Max.II$, is presented in the middle panel of Figure 1. The O'Connell effect of TIC 103656297 changes over time, as shown in the right panel of Figure 1. Late-type contact binary stars usually exhibit relatively strong magnetic activity. 
The noise of the TESS light curve, which is primarily centered $\sim 0.0012$ mag, was statistically analyzed by \citet{Ding+et+al+2024}. Under such a low noise level and considering the magnetic activity phenomenon of late-type stars, the $\Delta M_{ag}$ value clearly indicates the asymmetry of the light curve.

\begin{figure}[!ht]
\begin{center}

\begin{minipage}{5.5cm}
	\includegraphics[width=5.5cm]{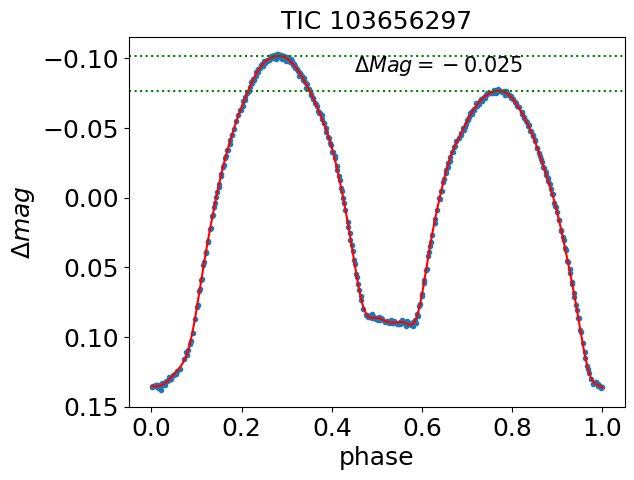}
\end{minipage}
\begin{minipage}{5.5cm}
	\includegraphics[width=5.5cm]{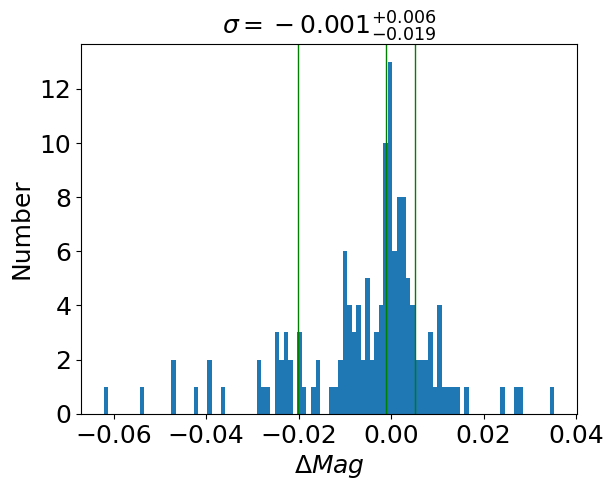}
\end{minipage}
\begin{minipage}{5.5cm}
	\includegraphics[width=5.5cm]{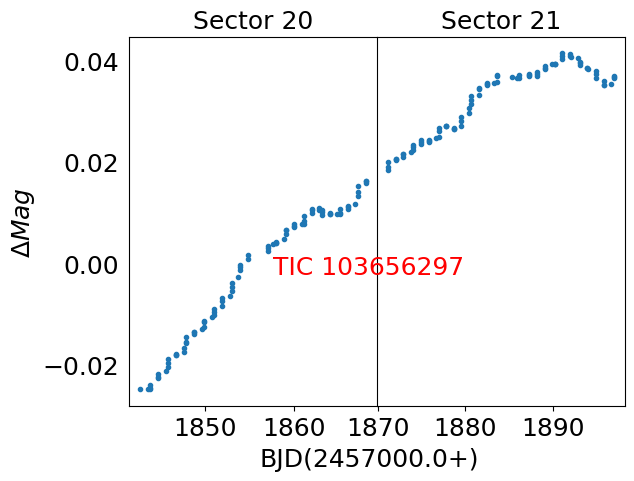}
\end{minipage}

\caption{Left: the blue dots depict the original light curve data, while the red line signifies the fitting outcome. The green lines highlight the Max.I and Max.II points within the light curves. Middle: the plot displays the $\Delta Mag$ distribution specifically for 143 light curves. Right: in Sectors 20 and 21, the O'Connell effect of TIC 103656297 varies with the passage of time.
}\label{Fig4}
\end{center}
\end{figure}

\section{Building a neural network model} 
In their analysis of the catalog referenced in \citet{Latkovi+et+al+2021}, \citet{Ding+et+al+2024} found that there were a total of 30 published targets recorded for the year 2020. In the light curve fitting of these targets, only BB Peg is particularly noteworthy as it features the addition of two cool spots. For the other targets, the fitting can be accomplished by adding just one cool spot. Therefore, the inclusion of a single cool spot can effectively fit the light curves of most targets. Therefore, we established a mapping relationship between the parameters and the light curves, where the parameters include those of the cool spots. In adherence to Occam's razor, our approach is to fit the light curve utilizing the minimum number of parameters. This stance, however, does not preclude the potential presence of a third light source, which necessitates an extended observational timeframe to scrutinize variations in the orbital period (O-C) \citep{Zhao+et+al+2021}. Consequently, our generated dataset omits any consideration of the third light.

\subsection{Sample data set}

In contact binary systems, the light curve is significantly affected by several key parameters, including the mass ratio ($q$), orbital inclination ($incl$), effective temperatures of the primary star ($T_1$) and the secondary star ($T_2$), fill-out factor ($f$), gravity-darkening coefficient ($g$), bolometric albedos ($A$), and the semi-major axis (sma), as well as the passband and parameters of any cool spots. The characteristics of a cool spot are typically defined by its colatitude ($colat$), longitude ($long$), angular radius ($radius$), and the temperature ratio compared to the local intrinsic value ($relteff$). To obtain accurate light curves for contact binaries, we use the Phoebe program, which relies on these fundamental parameters. The selected passband is TESS:T, which encompasses a broadband wavelength range from 600 to 1000 nm \citep{Ricker+et+al+2015}. To enhance the sample density, we adopt a random fraction method for generating these light curves.

The mass ratio ($q$) is distributed within the interval [0,1]. The orbital inclination ($incl$) ranges between 50 and 90 degrees. The fill-out factor ($f$) follows a distribution across the range [0,1]. The temperature of the primary star ($T_1$) is distributed within [4000K, 8000K]. The temperature ratio ($T_2/T_1$) is modeled by a Gaussian distribution with a mean ($\mu$) of 1 and a standard deviation ($\sigma$) of 0.2. The colatitude ($colat$) spans the interval [0,180]. The longitude ($long$) of the cool spot on the primary star is distributed throughout the range of 0 to 360 degrees. The angular radius ($radius$) ranges from 0 to 50 degrees, and the temperature ratio relative to the intrinsic local value ($relteff$) is distributed within [0.6,1]. We maintain Phoebe's default settings for limb-darkening, using ld$\_$func=‘interp’ for the linear interpolation algorithm, and for the reflection effect, we retain irrad$\_$method='none', corresponding to the 'Horvat' method, which is Lambertian. For each local point on a star, the limb-darkened intensity distribution is interpolated from pre-computed tables based on the effective temperature ($T_{eff}$), surface gravity ($\log g$), metallicity ([M/H]), and other relevant physical parameters \citep{prisa+et+al+2016}. For effective temperatures below 8000K, the gravity-darkening coefficient ($g$) is set to 0.32, and the bolometric albedo ($A$) is set to 0.6. The atmosphere model is selected as ck 2004. The semi-major axis (sma) is set to 1, enabling the acquisition of the relative radii of the primary and secondary stars.
The dataset comprises 200,000 samples.

{
\tiny
\begin{center}
\begin{longtable}{cccccccc}
\caption{Parameter Range of Contact Binaries.}\label{Table 1}\\
\hline\hline                          
Num     &parameter                   &range                              \\
\hline
\endhead
\hline

1    &$q$	                         &0-1                            \\
2    &$incl$($^{\circ}$)	         &50-90                           \\
3    &$T_1$(K)	                     &4000-8000 	                \\
4    &$T_2/T_1$                      &$\mu = 1$, $\sigma = 0.2$                      \\
5    &$f$                            &0-1                             \\
6    &$g_1=g_2$                      &0.32                                \\
7    &$A_1=A_2$                      &0.6                                \\
8    &passband                       & TESS:T                                \\
9    &sma                            & 1                                \\
10   & $colat$($^{\circ}$)           &0-180 \\
11   & $long$($^{\circ}$)            &0-360  \\
12   & $radius$($^{\circ}$)          &0-50 \\
13   & $relteff$                     &0.6-1\\
\hline
\end{longtable}
\end{center}
}

\subsection{A neural network model} 
Neural Network-Based Mapping of Parameters to Light Curves with Phase-Magnitude: A multilayer perceptron (MLP) neural network is utilized to establish the relationship between parameters and light curves, focusing specifically on the phase-magnitude characteristics. For this purpose, we developed a neural network model to aid in the analysis.

The configuration of the neural network we developed is depicted in Figure 2. This architecture primarily comprises an input layer, multiple hidden layers, and an output layer. The values shown in Figure 2 correspond to the number of nodes in each of these layers. The input layer of the model contains 9 nodes, reflecting the 9 input parameters ($T_1, q, incl, f, T_2/T_1, colat, long, radius, relteff$). Activation functions are incorporated into both the input and hidden layers, enabling the resolution of non-linear relationships that cannot be expressed linearly. We opted for the ReLU activation function \citep{He+et+al+2015} for both models. The loss function measures the discrepancy between the predicted and actual values of the model. The variation of the loss function on the validation set is used to select an optimal model for further training on the training set. The Mean Square Error (MSE) is chosen as the loss function. The Adam optimizer \citep{Kingma+et+al+2014} was selected for the neural network. The model requires approximately 0.01 seconds to generate a light curve with 100 points when executed on a computer equipped with 32G of random-access memory, an i7-8700 CPU operating at 3.20 GHz. Under identical conditions, the Phoebe program takes approximately 5.72 seconds to generate a light curve with 100 points. The speed at which the neural network model generates light curves is two orders of magnitude faster than that of Phoebe. The i7-8700 CPU has 6 cores and can support 12 threads as it allows two threads per core. Both the neural network model and Phoebe were run in single-thread mode when generating light curves based on parameters for the speed comparison. No parallel mode using multiple threads was employed.

\begin{figure}[!ht]
\begin{center}

\begin{minipage}{16cm}
	\includegraphics[width=16cm]{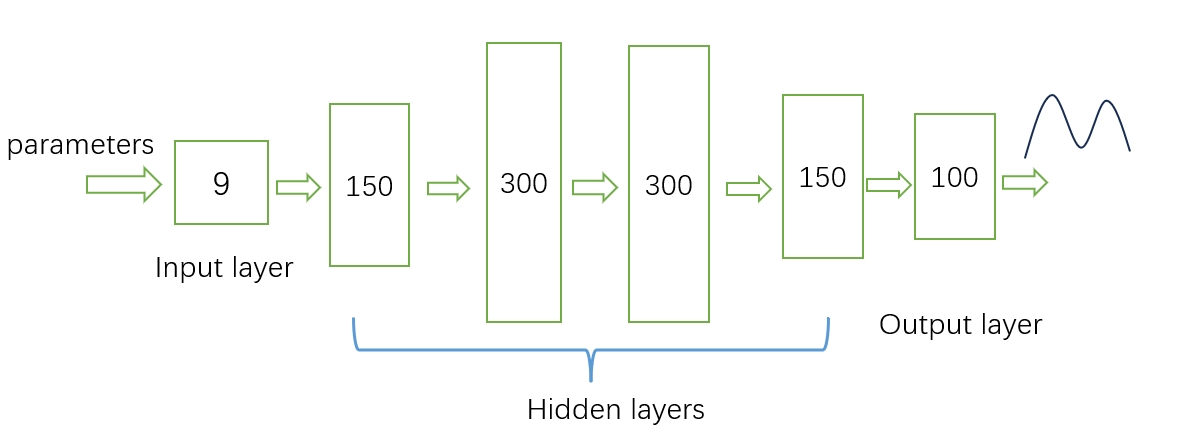}
\end{minipage}

\caption{This figure illustrates the structure of the neural network, where the node number represents the count of neurons.}\label{Fig1}
\end{center}
\end{figure}

\vspace{2em}
\subsection{Precision of light curves generated by neural network model}
The original dataset is partitioned into three distinct subsets following an 8:1:1 distribution: training, validation, and test sets. The training set is utilized to construct a neural network model, which undergoes a learning process to optimize its parameters. Once the training phase is complete, the resulting model is considered the final iteration. Its effectiveness is then verified through evaluation against the validation set. Finally, the precision with which the model generates light curves is measured by testing it against the independent test set.

The precision of light curves generated by the model is evaluated by examining the standard deviation of the residuals. This assessment involves comparing the distribution of residuals from the synthetic light curves produced by Phoebe with those generated by the model. As depicted in the left panel of Figure 3, the blue light curve was generated by Phoebe using the following input parameters: $T_1 = 7342$ K, $incl = 73.57^{\circ}$, $q = 0.27$, $f = 0.91$, $T_2/T_1 = 1.01$, $relteff = 0.79$, $colat=28^{\circ}$, $long=-95^{\circ}$, $radius=11^{\circ}$. The model, employing the same set of parameters, produced the orange light curve in the same panel. The standard deviation of the residuals between these two curves is calculated to be 0.0005. Additionally, we computed the standard deviation of residuals across a sample of 20,000 light curves. The right panel of Figure 3 illustrates that the standard deviations of the residuals are predominantly distributed around $0.0006^{+0.0003}_{-0.0002}$. While the number of hidden layers and neurons in this work may not be optimal, they suffice to meet the requirements in terms of generating light curve accuracy.

\begin{figure}[!ht]
\begin{center}

\begin{minipage}{8cm}
	\includegraphics[width=8cm]{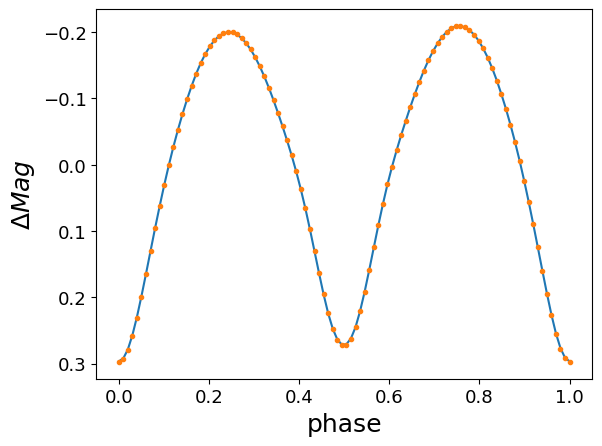}
\end{minipage}
\begin{minipage}{8cm}
	\includegraphics[width=8cm]{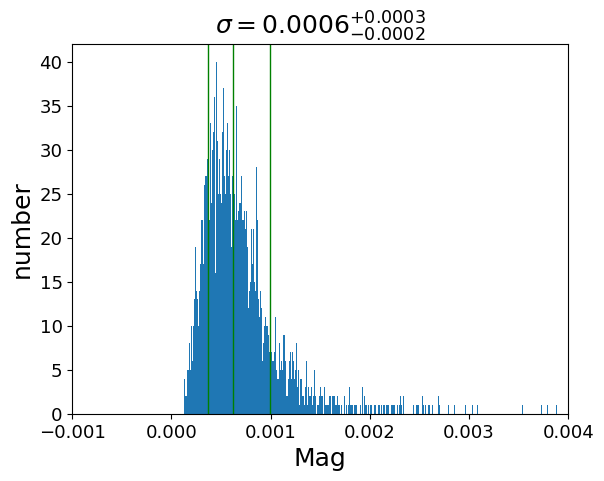}
\end{minipage}

\caption{Left: The blue light curve was synthesized by Phoebe, whereas the orange light curve was generated by the model employing the same set of input parameters. Right: The distribution of the standard deviation of the residuals between the predicted light curves produced by the model and the synthetic light curves produced by Phoebe is depicted.}\label{Fig2}
\end{center}
\end{figure}

\clearpage
\section{TESS Survey: Contact Binary Parameters}
Employing a neural network (NN$_{nol3}$) model that has been trained and fine-tuned, in conjunction with the Markov Chain Monte Carlo (MCMC) algorithm, we successfully derive the posterior distributions of the parameters characterizing contact binaries. We then further estimated the absolute parameters of these targets.

\subsection{Relative parameters}
The relative parameters for 143 contact binaries are determined through the utilization of a neural network (NN$_{nol3}$) model in conjunction with the Markov Chain Monte Carlo (MCMC) algorithm. An illustrative case is presented using TIC 164720673 to demonstrate the extraction of these parameters for the specific object. Subsequently, the fitting outcomes for four objectives are detailed.

Initially, we employed the NN$_{nol3}$ model along with the MCMC algorithm to derive the posterior distribution of the parameters for the target identified as TIC 164720673. A total of 30 parameter space walkers were utilized. To ascertain the convergence of the MCMC chain, multiple criteria have been established to select an adequate chain length, such as a duration that is 10–20 times the integrated autocorrelation time \citep{Conroy+et+al+2020, Li+et+al+2021}. For contact binaries exhibiting flat-bottom minima in their light curves, a significant proportion are identified as low mass ratio systems. In such systems, the primary component (the hotter star) plays a dominant role in determining the system's effective temperature, $T_{sys}$. Hence, for contact binaries with low mass ratios, where the temperature of the primary component largely dictates the system temperature, we consider $T_{sys}$ to be approximately equal to $T_1$ (the effective temperature of the primary star). The effective temperature of the primary star ($T_1)$ was set at 6294 K by cross-referencing with Gaia DR2 data \citep{Gaia+et+al+2018}. We opted for a chain length greater than 40 times the integrated autocorrelation time to guarantee convergence. The light curve underwent 150,000 iterations of the MCMC parameter search post an initial 5000 steps. The final 2000 steps were saved and a corner plot was generated. The posterior distribution of the parameters ($incl, q, f, T_2/T_1, relteff, colat, long, radius$) is showcased in Figure 4. Figure 5 depicts the original light curve using blue dots. The red line in the Figure 5 represents the light curve produced by the model using the derived parameters. The x points on the same panel indicate the light curve generated by Phoebe with identical parameters. The goodness of fit metric ($R^2$) between the Phoebe-generated light curve using these parameters ($Teff, q, incl, f, T_2/T_1, relteff, colat, long, radius$) and the original light curve stands at 0.989.

The NN$_{nol3}$ model and MCMC algorithm were applied to fit four noteworthy objectives labeled TIC 6171564, TIC 89428764, TIC 93053299, and TIC 97328396. The goodness of fit ($R^2$) for each of these targets exceeds 0.99. Table 2 outlines the parameters for four contact binaries, while the fitting results for these targets are depicted in Figure 6.

\begin{figure}[!ht]
\begin{center}
\begin{minipage}{15cm}
	\includegraphics[width=15cm]{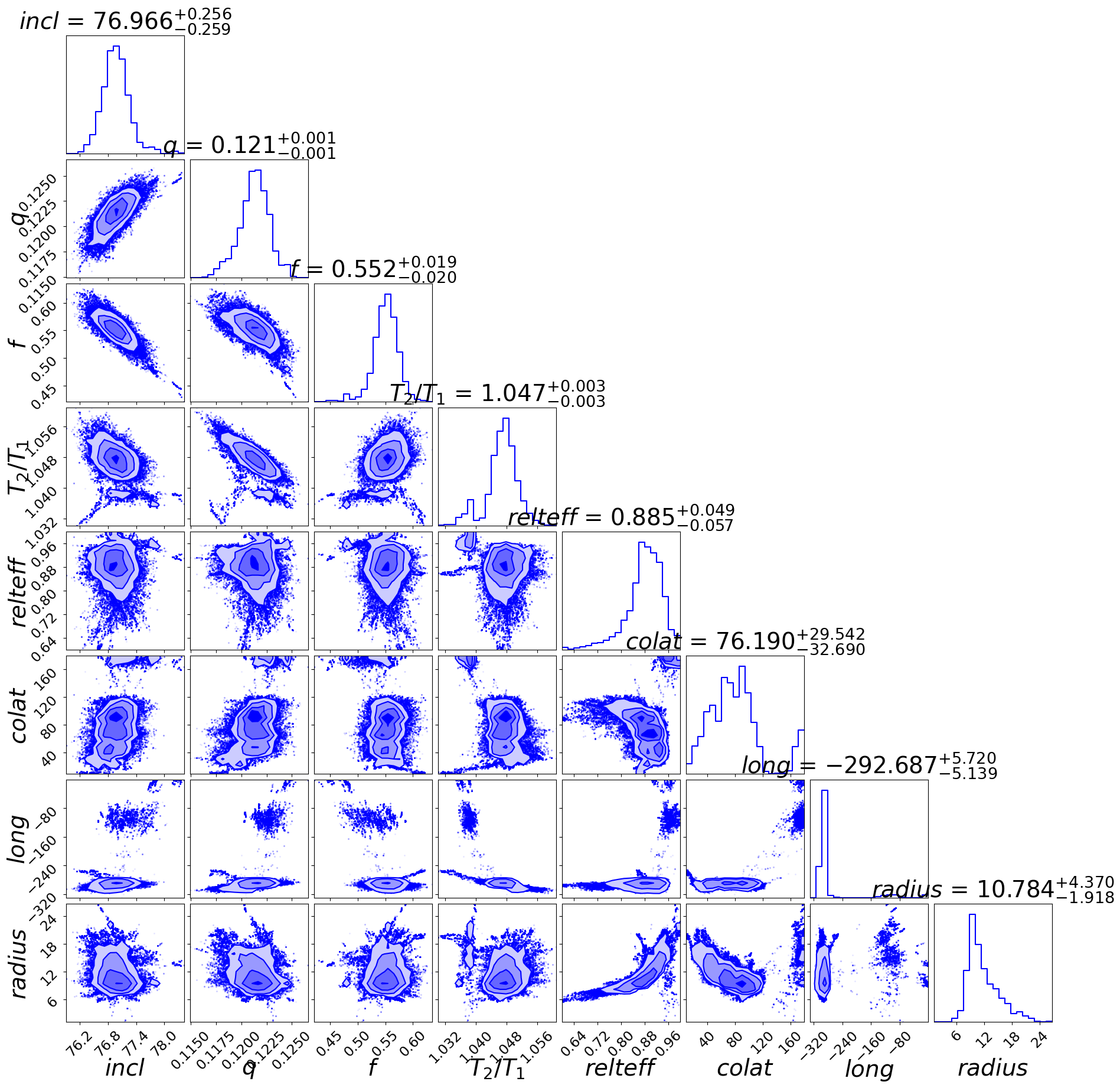}
\end{minipage}

\caption{The posterior parameter distributions for TIC 164720673 are presented.}\label{Fig5}
\end{center}
\end{figure}

\begin{figure}[!ht]
\begin{center}
\begin{minipage}{10cm}
	\includegraphics[width=10cm]{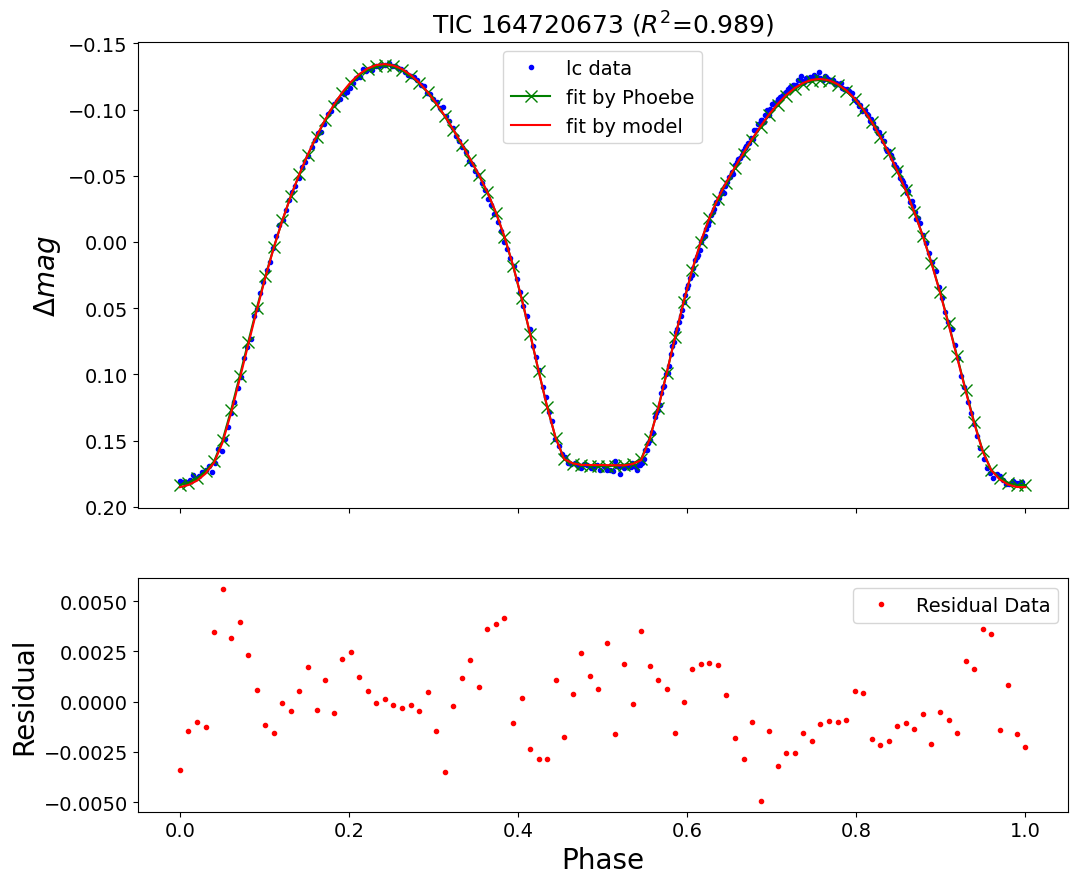}
\end{minipage}
\caption{Fitting Results for TIC 164720673 are presented.}\label{Fig6}
\end{center}
\end{figure}

{
\tiny
\begin{center}
\begin{longtable}{ccccccccccccccc}
\caption{The essential parameters of the four contact binary systems}\label{Table 2}\\
\hline\hline                          
name     &$T_1$(K)   &$incl(\circ)$  &$\sigma_{incl}$  &$q$  &$\sigma_{q}$  &$f$  &$\sigma_{f}$   &$\frac{T_2}{T_1}$  &$\sigma_{\frac{T_2}{T_1}}$  &$R^2$                   \\
\hline
\endhead
\hline

TIC 671564 & 5607 & 81.637 & 0.443 & 0.160 & 0.002 & 0.668 & 0.031 & 1.055 & 0.006 & 0.995 \\   
TIC 89428764 & 6541 & 76.439 & 0.257 & 0.136 & 0.001 & 0.318 & 0.022 & 0.995 & 0.002 & 0.999 \\ 
TIC 93053299 & 5954 & 76.706 & 0.494 & 0.126 & 0.002 & 0.525 & 0.042 & 1.020 & 0.003 & 0.997 \\
TIC 97328396 & 5935 & 86.344 & 0.469 & 0.291 & 0.002 & 0.502 & 0.010 & 1.016 & 0.002 & 0.993 \\

\hline
\end{longtable}
\end{center}
}

\begin{figure}[!ht]
\begin{center}
\begin{minipage}{7.2cm}
	\includegraphics[width=7.2cm]{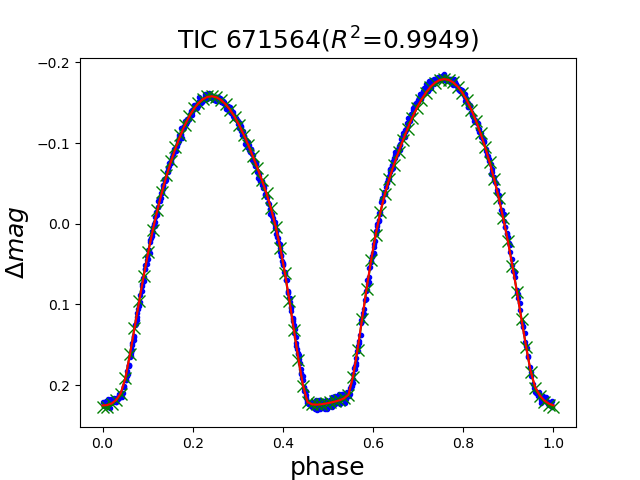}
\end{minipage}
\begin{minipage}{7.2cm}
	\includegraphics[width=7.2cm]{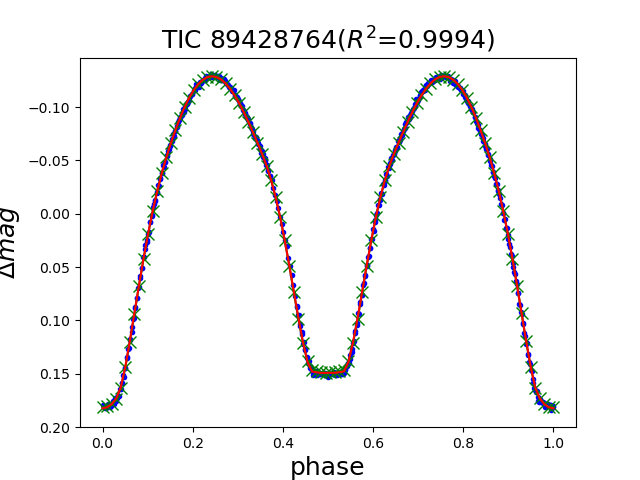}
\end{minipage}
\begin{minipage}{7.2cm}
	\includegraphics[width=7.2cm]{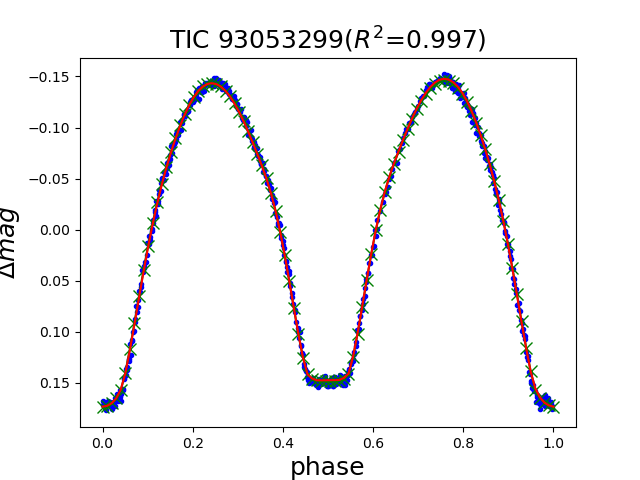}
\end{minipage}
\begin{minipage}{7.2cm}
	\includegraphics[width=7.2cm]{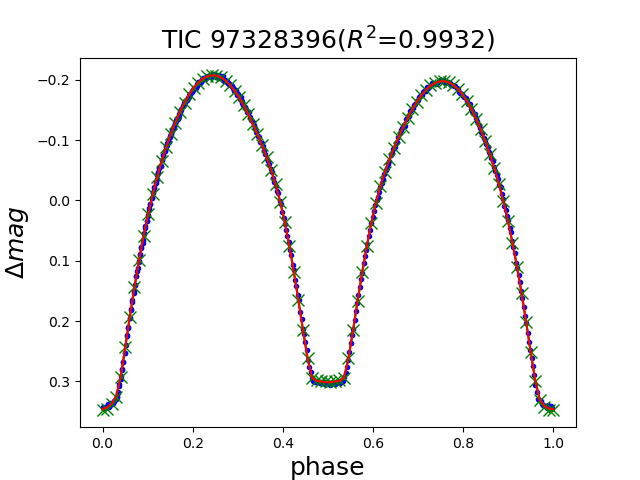}
\end{minipage}

\caption{The fitting results for four targets are depicted, with the blue dots indicating the original light curves. The red lines represent the light curves generated by the trained model, and the green x dots denote the light curves produced by Phoebe.}\label{Fig7}
\end{center}
\end{figure}

The catalog presented in Table 3 \footnote{https://github.com/dingxu6207/Data\label{fn:github}} (Columns 1-23) encapsulates the relative parameters and associated uncertainties for 143 contact binaries. Initially, the catalog furnishes foundational data about each target, including target name, orbital period, period uncertainty, and the primary star's effective temperature ($T_1$) as derived from Gaia DR2. Subsequently, parameters inferred through neural network (NN$_{nol3}$) model and Markov Chain Monte Carlo (MCMC) simulations are included. These parameters encompass the orbital inclination ($incl$), mass ratio ($q$), temperature ratio ($T_2/T_1$), fill-out factor ($f$), colatitude ($colat$), longitude ($long$), and angular radius ($radius$), alongside their respective errors. The catalog also features the temperature ratio relative to the local intrinsic value ($relteff$) and its corresponding error.

Using the derived parameters, the Phoebe software computes the relative radii of the primary ($r_1$) and secondary ($r_2$) stars. This is the equivalent radius in terms of volume. First, calculate the volume of the non-sphere. Then, assume that the equivalent radius refers to the radius of a sphere having the same volume as the non-sphere. The relative radii $r_1$ and $r_2$ are the ratios of the primary star's radii to the semi-major axis of the orbit and the secondary star's radii to the semi-major axis of the orbit, respectively. Additionally, the catalog provides the goodness of fit metric ($R^2$), which quantifies the alignment between the Phoebe-generated light curve and the original observed light curve.

\clearpage
{
\tiny
\begin{center}
\begin{longtable}{cccc}
\caption{Catalog Listing: Fundamental Parameters of 143 Contact Binaries}\label{Table 3}\\
\hline\hline                          
Num     &Column          &Units           &Explanations                     \\
\hline
\endhead
\hline

1    &name	                &                     &TESS object name          \\
2    &$T_1$ 	            &K                    &Effective temperature of primary star (from Gaia DR2)    \\
3    &Period                &day                  &Orbital period              \\
4    &$\sigma_{Period}$     &day                  &Uncertainty in Orbital period              \\
5    &incl 	                &$^{\circ}$     	  &Orbital inclination           \\
6    &$\sigma_{incl}$	    &$^{\circ}$	          &Uncertainty in incl     \\
7   &$q$                     &                     &mass ratio    \\
8   &$\sigma_{q}$	        &                     &Uncertainty in $q$     \\
9   &$T_2/T_1$             &                     &Temperature ratio    \\
10   &$\sigma_{T_2/T_1}$    &                     &Uncertainty in $T_2/T_1$  \\ 
11   &$f$                     &                     &Fill-out factor  \\
12   &$\sigma_{f}$          &                     &Uncertainty in $f$  \\
13   &$relteff$                 &                   &Temperature ratio (cool spot) \\
14   &$\sigma_{relteff}$        &                     &Uncertainty in $relteff$  \\
15   &$colat$                 &                   &colatitude (cool spot) \\
16   &$\sigma_{colat}$        &                     &Uncertainty in $colat$  \\
17   &$long$                 &                   &longitude(cool spot) \\
18   &$\sigma_{long}$        &                     &Uncertainty in $long$  \\
19   &$radius$                 &                   &angular radius (cool spot) \\
20   &$\sigma_{radius}$     &                     &Uncertainty in $radius$  \\
21   &$r_{1}$               &                     &Relative radii of primary star  \\
22   &$r_{2}$               &                     &Relative radii of secondary star  \\
23   &$R^2$                 &                    &Goodness of fit  \\  
24    &$a$     	            &$R_{\odot}$         &semi-major axis   \\ 
25    &$\sigma_a$     	    &$R_{\odot}$         &Uncertainty in $a$   \\ 
26    &$M_1$ 	            &$M_{\odot}$         &mass of primary star    \\
27    &$\sigma_{M_1}$ 	    &$M_{\odot}$         &Uncertainty in $M_1$    \\ 
28    &$M_2$ 	            &$M_{\odot}$         &mass of secondary star    \\
29    &$\sigma_{M_2}$ 	    &$M_{\odot}$        &Uncertainty in $M_2$    \\ 
30    &$R_1$ 	            &$R_{\odot}$        &radii of primary star           \\
31    &$\sigma_{R_1}$ 	    &$R_{\odot}$        &Uncertainty in $R_1$           \\
32    &$R_2$	            &$R_{\odot}$	    &radii of secondary star     \\
33    &$\sigma_{R_2}$	    &$R_{\odot}$	    &Uncertainty in $R_2$      \\
34   &$L_1$                 &$L_{\odot}$        &luminosity of primary star    \\
35   &$\sigma_{L_1}$        &$L_{\odot}$        &Uncertainty in $L_1$    \\
36   &$L_2$ 	            &$L_{\odot}$        &luminosity of secondary star     \\
37   &$\sigma_{L_2}$ 	    &$L_{\odot}$        &Uncertainty in $L_2$     \\
\hline
\end{longtable}
\end{center}
}

\clearpage
\subsection{Absolute parameters}
In the analysis of the catalog referenced in \citet{Latkovi+et+al+2021}, a total of 153 contact binaries were identified, for which absolute parameters were determined through the combined assessment of their radial velocity and photometric light curves. Notably, all of these contact binaries have temperatures less than 10,000 K. Our findings indicate a strong correlation between the orbital period ($P$) and the semi-major axis ($a$), as illustrated in Figure 7. Only one target deviates significantly, named HV UMa \citep{Csak+et+al+2000}, with a period of 0.35 days and a semi-major axis of 5.002 $R_{\odot}$.

Random Sample Consensus (RANSAC) is a robust iterative algorithm designed to estimate mathematical models from data sets containing outliers \citep{Dao+et+al+2024} . By systematically sampling subsets of the data and identifying consensus within these subsets, RANSAC effectively distinguishes between inliers and outliers, thereby enabling a more accurate model estimation. This method is particularly valuable in scientific research, where data quality can be compromised by noise and erroneous measurements, ensuring that the derived models remain reliable and representative of the true underlying relationships. We employed the RANSAC algorithm to remove some outliers and fit an inlier dataset, as shown in Figure 7.
The connection between $P$ and a can be expressed using the subsequent linear formula,

\begin{equation}
    a = 0.476(\pm 0.053) + 5.687(\pm 0.117) \times P  
\end{equation}

The values of the semi-major axis ($a$) were calculated based on the orbital period ($P$). Their respective radii ($R_1,R_2$) are determined by the semi-major axis ($a$) and the relative radii ($r$).

\begin{equation}
    R = a \times r
\end{equation}

The masses of two components were determined using mass ratio ($q$) and Kepler's third law. $G$ stands for gravitational constant, and $P$ represents the orbital period of system.
\begin{equation}
\frac{a^3}{G(M_1+M_2)} = \frac{P^2}{4 \pi ^2}
\end{equation}

The Stefan-Boltzmann law was employed to calculate the luminosity of the primary star ($L_1$), followed by the determination of the secondary star’s luminosity ($L_2$).
\begin{equation}
    L_1 = R_1^2 \times (4\pi \sigma T_1^4)
\end{equation}

\begin{equation}
    L_2 = R_2^2 \times (4\pi \sigma T_2^4)
\end{equation}

The calculated parameters include the semi-major axis ($a$), mass ($M_1,M_2$), radii ($R_1,R_2$), and luminosity ($L_1,L_2$), as presented in Table 3 (Columns 24-37).

\begin{figure}[!ht]
\begin{center}
\begin{minipage}{10cm}
	\includegraphics[width=10cm]{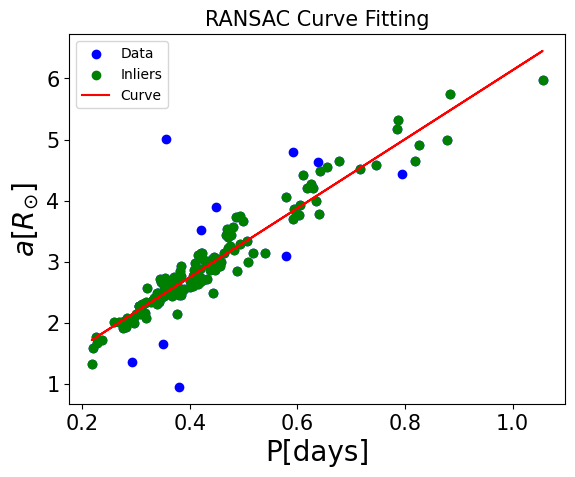}
\end{minipage}
\caption{The correlation between the orbital period ($P$) and the semi-major axis ($a$) for contact binary systems with temperatures below 10,000 K, as derived from \citet{Latkovi+et+al+2021}.}\label{Fig8}
\end{center}
\end{figure}

The total luminosity ($L_{total} = L_1 + L_2$) obtained in this study was compared with both the luminosity from Gaia DR2 and the one derived from the period–color–magnitude relation. We conduct name matching via Simbad \footnote{https://simbad.cds.unistra.fr/simbad/} and retrieve the parameters from the Gaia DR2 catalog. These parameters encompass parallax ($Plx$), apparent magnitudes ($G, Bp, Rp$), proper motions ($pmRA, pmDE$), effective temperature ($Teff$), and luminosity ($Lum$). Subsequently, we present a catalog of the relevant Gaia DR2 parameters \textsuperscript{\ref{fn:github}}. The period–color–magnitude relation developed and utilized by Rucinski and several collaborators \citep{Rucinski+et+al+1994, Rucinski+et+al+1995, Rucinski+et+al+1997, Rucinski+et+al+2006, Mateo+et+al+2017} serves as a highly effective method for estimating the absolute magnitude of W UMa contact binaries. \citet{Mateo+et+al+2017} have provided a dataset of $M_v = M_v(logP)$ calibrations for 318 W UMa-type (EW)
contact binaries derived from Gaia data. Consequently, initially we can obtain the absolute magnitudes using the formula presented by \citet{Mateo+et+al+2017} as 

\begin{equation}
    M_V = 3.73-8.67 \times (logP+0.4), (0.275 < P < 0.575 \quad days)   
\end{equation}

\begin{equation}
    M_V = 3.28-7.54 \times (logP+0.35)+9.96 \times (logP+0.35)^2, (0.22 < P < 0.9 \quad days)
\end{equation}

Then we can obtain the total luminosity with the relationships as
\begin{equation}
    L_{total} = L_1+L_2 =  10^{-0.4(M_V+BC_v-4.73)} \times L_{\odot}
\end{equation}

The bolometric correction ($BC_v$) is derived using the dataset from \citet{Pecaut+2013}. The comparison results of the total luminosity (\( L_{\text{total}} \)) calculations are presented in Figure 8. It can be seen from Figure 8 that the obtained results are linearly correlated, which proves that the method for solving the parameters in this paper is also feasible. 
The dispersion, when compared with the luminosity from Gaia DR2 as shown in the left panel of Figure 8, is relatively small. The main reason is that the relationship between the semi-major axis ($a$) and the orbital period ($P$) is calculated using an empirical formula. The empirical formula shows dispersion, as shown in Figure 7. The dispersion in the right panel of Figure 8 is larger because both the relationship between the semi-major axis ($a$) and the orbital period ($P$) and the relationship between the absolute magnitude ($M_v$) and the orbital period ($P$) are calculated using empirical formulas. Among the 143 targets, 88 have periods satisfying the range \(0.275 < P < 0.575 \, \text{days}\), and 129 meet the criteria of \(0.22 < P < 0.9 \, \text{days}\). The comparison results of the \( L_{\text{total}} \) calculations are presented in Figure 8, with the deviations being $\sigma_{log(L_{total})} = 0.165 $ and $\sigma_{log(L_{total})} = 0.175 $, respectively.

\begin{figure}[!ht]
\begin{center}
\begin{minipage}{8cm}
	\includegraphics[width=8cm]{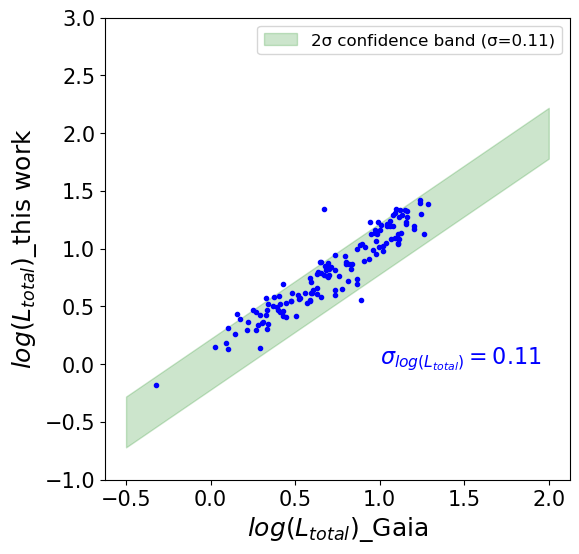}
\end{minipage}
\begin{minipage}{8cm}
	\includegraphics[width=8cm]{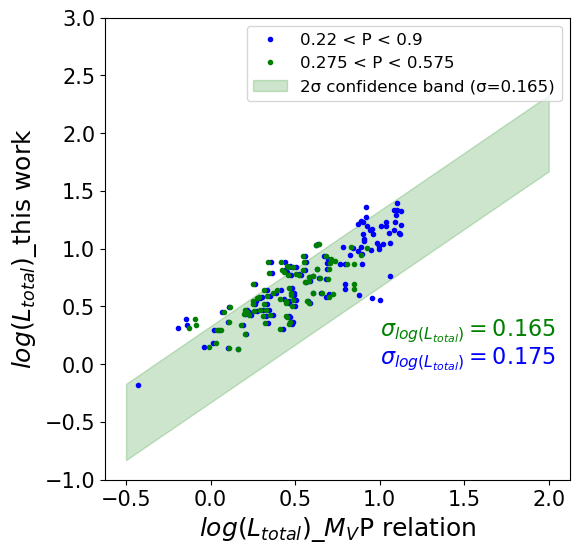}
\end{minipage}
\caption{The comparison results of the total luminosity (\( L_{\text{total}} \)) calculations are presented. The left-hand figure shows the comparison with the luminosity provided by Gaia DR2, while the right-hand figure shows the comparison with the one derived from the period–color–magnitude relation. The blue dots and the light-green dots represent the use of different empirical formulas based on the period.
}\label{Fig84}
\end{center}
\end{figure}

\section{Discussion}
\subsection{Compare another catalog }

Through a thorough review of pertinent literature, \citet{Latkovi+et+al+2021} performed a statistical and analytical examination of 700 W UMa stars that had been individually analyzed. Our dataset includes 32 entries that correspond with the catalog compiled by \citet{Latkovi+et+al+2021}, sourced through SIMBAD. We undertook a comprehensive comparison of all 32 overlapping individual objects. The key parameters for these 32 targets are detailed in Table 4. Figure 9 provides a visual comparison of the results showcased in Table 4. The standard deviation of the residuals for the primary star's effective temperature ($T_1$) across these 32 targets is determined to be 499 K. Similarly, the standard deviation for the orbital inclinations ($incl$) among the same set of targets is found to be 4.05 degrees. For the mass ratios ($q$), the standard deviation of the residuals is 0.02, while for the fill-out factors ($f$), it is 0.13. Furthermore, the standard deviation of the residuals for the temperature ratios ($T_2/T_1$) of these 32 targets is determined to be 0.04. Previous research by \citet{Pribulla+et+al+2003} has demonstrated that the mass ratios of total eclipsing binaries can be accurately assessed without the necessity of spectroscopic observations. As a result, the mass ratios for such total eclipsing binaries exhibit a notably high level of precision.

\begin{table*}
    \centering
	\caption{Parameter Comparison: $T_1, incl, q, f, T_2/T_1$ results from this study, $T_1^*,incl^*, q^*, f^*, T_2/T_1^*$ results from \citet{Latkovi+et+al+2021}.}\label{Table 5}
	\begin{tabular}{ccccccccccc} 
\hline \hline                        
 name &  $T_1$ & $incl$ & $q$ & $f$ &  $T_2/T_1$ & $T_1^*$  &$incl^*$ & $q^*$ & $f^*$ &   $T_2/T_1^*$    \\
\hline
TIC 144376560 & 7135 & 87.820 & 0.086 & 0.570 & 1.082 & 7410 & 81.100 & 0.076 & 0.487 & 0.994 \\
 TIC 271495349 & 6688 & 86.640 & 0.430 & 0.375 & 1.001 & 6250 & 87.040 & 0.452 & 0.385 & 0.991 \\
 TIC 317470793 & 5673 & 86.620 & 0.332 & 0.063 & 0.991 & 6440 & 85.100 & 0.332 & 0.133 & 1.000 \\
 TIC 170676440 & 6907 & 85.530 & 0.229 & 0.289 & 0.870 & 6881 & 69.400 & 0.300 & 0.691 & 0.739 \\
 TIC 296657504 & 5769 & 85.260 & 0.271 & 0.183 & 1.030 & 5721 & 83.300 & 0.257 & 0.343 & 1.066 \\
 TIC 274967957 & 5774 & 84.900 & 0.261 & 0.123 & 0.971 & 6580 & 87.310 & 0.296 & 0.464 & 0.952 \\
 TIC 417052182 & 5899 & 84.870 & 0.118 & 0.484 & 0.994 & 6215 & 80.760 & 0.107 & 0.577 & 1.026 \\
 TIC 5674169 & 5874 & 84.760 & 0.199 & 0.473 & 1.030 & 5940 & 81.900 & 0.190 & 0.494 & 0.995 \\
 TIC 343409837 & 6101 & 83.960 & 0.105 & 0.486 & 1.028 & 7035 & 85.200 & 0.115 & 0.391 & 0.950 \\
 TIC 157175641 & 6138 & 83.580 & 0.319 & 0.218 & 1.027 & 6014 & 82.070 & 0.299 & 0.298 & 1.021 \\
 TIC 233636431 & 6896 & 83.460 & 0.244 & 0.249 & 0.997 & 6705 & 81.990 & 0.248 & 0.347 & 0.976 \\
 TIC 265015162 & 5787 & 83.420 & 0.118 & 0.669 & 1.070 & 6200 & 82.600 & 0.104 & 0.698 & 1.051 \\
 TIC 359987981 & 5856 & 82.510 & 0.084 & 0.550 & 0.997 & 6640 & 85.100 & 0.100 & 0.446 & 0.938 \\
 TIC 456797644 & 7019 & 82.480 & 0.193 & 0.728 & 0.994 & 6900 & 83.640 & 0.250 & 0.843 & 1.002 \\
 TIC 267043786 & 6720 & 82.450 & 0.304 & 0.213 & 0.967 & 6900 & 75.890 & 0.354 & 0.366 & 1.021 \\
 TIC 294273900 & 6028 & 82.310 & 0.214 & 0.315 & 1.047 & 6383 & 81.800 & 0.220 & 0.378 & 0.993 \\
 TIC 290034099 & 7810 & 82.250 & 0.319 & 0.402 & 0.980 & 6820 & 81.800 & 0.306 & 0.469 & 0.970 \\
 TIC 17563811 & 6277 & 82.220 & 0.099 & 0.638 & 1.037 & 6700 & 82.100 & 0.101 & 0.754 & 1.016 \\
 TIC 372127422 & 7218 & 82.160 & 0.186 & 0.580 & 0.991 & 6900 & 86.390 & 0.220 & 0.693 & 1.010 \\
 TIC 57297550 & 7038 & 81.820 & 0.069 & 0.931 & 1.023 & 8300 & 85.430 & 0.065 & 0.837 & 1.026 \\
 TIC 250203533 & 5733 & 80.940 & 0.167 & 0.438 & 1.028 & 6460 & 79.190 & 0.145 & 0.568 & 1.032 \\
 TIC 411451519 & 7035 & 80.900 & 0.218 & 0.289 & 0.982 & 6912 & 81.000 & 0.210 & 0.434 & 0.997 \\
 TIC 29287800 & 6125 & 80.540 & 0.254 & 0.287 & 1.033 & 6100 & 79.500 & 0.232 & 0.228 & 1.065 \\
 TIC 219738202 & 6904 & 79.620 & 0.265 & 0.539 & 0.972 & 6980 & 81.800 & 0.288 & 0.604 & 0.968 \\
 TIC 458490358 & 6529 & 79.200 & 0.178 & 0.495 & 1.020 & 6500 & 78.400 & 0.172 & 0.592 & 1.000 \\
 TIC 323442776 & 6116 & 79.190 & 0.287 & 0.282 & 0.984 & 6605 & 78.490 & 0.275 & 0.268 & 0.995 \\
 TIC 55753802 & 6610 & 78.980 & 0.198 & 0.510 & 1.006 & 6200 & 78.600 & 0.206 & 0.592 & 0.998 \\
 TIC 13070701 & 7211 & 77.770 & 0.104 & 0.419 & 1.000 & 6347 & 83.300 & 0.133 & 0.535 & 0.934 \\
 TIC 146520491 & 6484 & 75.160 & 0.135 & 0.330 & 1.001 & 6008 & 73.800 & 0.120 & 0.627 & 0.983 \\
 TIC 11480757 & 7002 & 75.100 & 0.104 & 0.329 & 1.007 & 6725 & 68.710 & 0.111 & 0.573 & 0.920 \\
 TIC 103656297 & 5886 & 75.100 & 0.070 & 0.178 & 1.041 & 6100 & 76.290 & 0.078 & 0.008 & 1.000 \\
 TIC 14545690 & 5883 & 70.960 & 0.080 & 0.407 & 1.106 & 5960 & 76.580 & 0.106 & 0.776 & 1.020 \\	             
\hline
\end{tabular}	
\end{table*}

\begin{figure}[!ht]
\begin{center}
\begin{minipage}{16cm}
	\includegraphics[width=16cm]{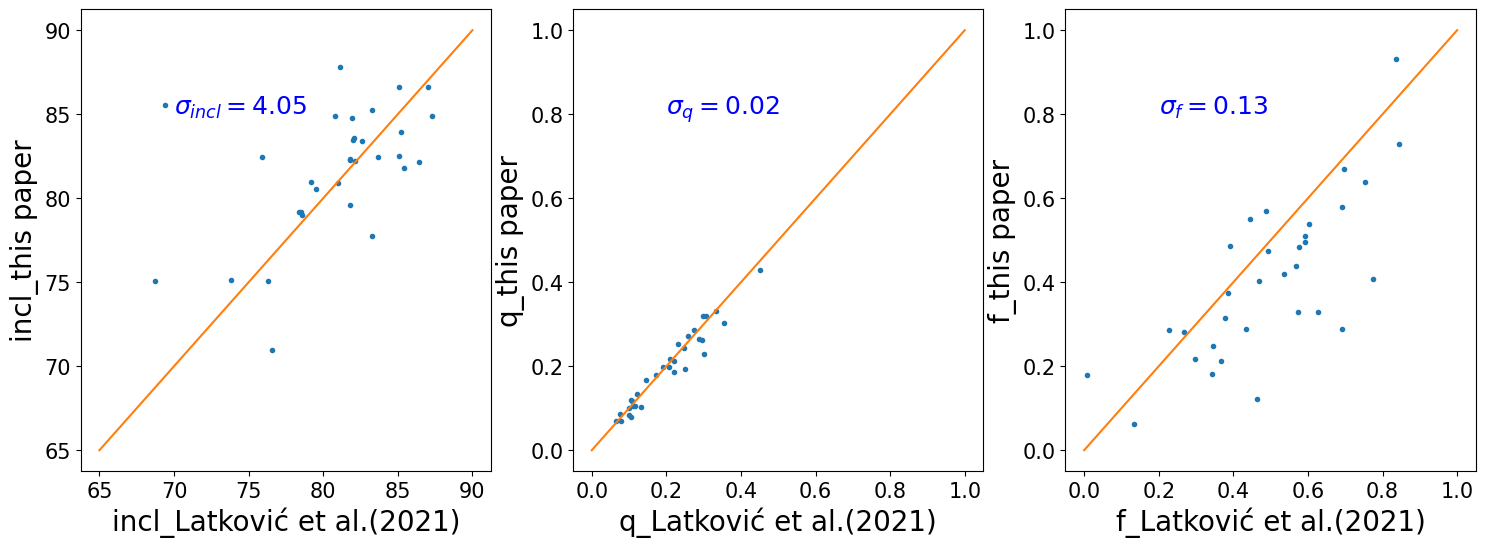}
\end{minipage}
\caption{A comparative analysis of the orbital inclination ($incl$), mass ratio ($q$), and fill-out factor ($f$) as presented in Table 4. The orange line is the 1:1 line.}\label{Fig9}
\end{center}
\end{figure}

For contact binaries displaying flat-bottom minima in their light curves, a notable proportion are classified as systems with low mass ratios.
The luminosity of the secondary star is lower, and the spectrum of the system is essentially characterized by the primary star’s spectral features.
To better illustrate this, we compared the temperatures derived in this study with those obtained from a low-resolution spectroscopic survey conducted using the Large Sky Area Multi-Object Fiber Spectroscopic Telescope (LAMOST) \citep{Luo+et+al+2015}. We cross-referenced 143 targets with the LAMOST DR11 low-resolution spectra, and LAMOST provided spectroscopic temperature data for 26 of these targets. The standard deviation of the residuals for the effective temperature of the primary star ($T_1$) across these 26 targets is found to be 376 K in Figure 10.

\begin{figure}[!ht]
\begin{center}
\begin{minipage}{7cm}
	\includegraphics[width=7cm]{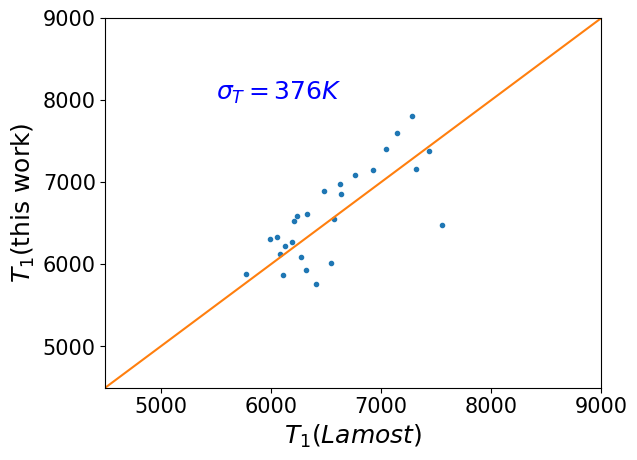}
\end{minipage}
\caption{The effective temperatures of the primary star obtained from LAMOST and those derived in this study. The orange line is the 1:1 line.}\label{Fig81}
\end{center}
\end{figure}

\subsection{New low mass ratio contact binary candidates }
\citet{Christopoulou+et+al+2022} detail the identification and photometric analysis of 30 newly discovered low mass ratio (LMR) totally eclipsing contact binaries with mass ratios $q \leq 0.25$, identified within the data from the Catalina Sky Survey. In this work, \citet{Christopoulou+et+al+2022} also listed 178 known low mass ratio contact binaries.
\citet{Lalounta+et+al+2024} report the discovery of 7 additional totally eclipsing low mass ratio (LMR) systems derived from the Catalina Sky Surveys (CSS). There are a total of 215 low mass ratio contact binary targets. In our work, we provide a catalog of parameters for 143 totally eclipsing contact binaries, of which 115 have mass ratios $q \leq 0.25$. When we cross-referenced these 115 targets with the previously mentioned 215 objects by SIMBAD, we identified 96 targets not included in the existing catalog, as detailed in Table 5. These 96 objects are likely to represent new low mass ratio contact binary candidates.

\begin{table*}
    \centering
	\caption{List of 96 Newly Identified Low Mass Ratio Contact Binary Candidates from this study}\label{Table 6}
	\begin{tabular}{cccccccc} 
\hline \hline                        
name &  $q$ &     name &     $q$ &  name &   $q$    \\
\hline
TIC 103656297 & 0.070 & TIC 114500611 & 0.152 & TIC 11480757 & 0.104 \\
TIC 115105140 & 0.164 & TIC 115906237 & 0.129 & TIC 116163356 & 0.151 \\
TIC 123038007 & 0.149 & TIC 138256097 & 0.232 & TIC 140757590 & 0.182 \\
TIC 141871560 & 0.168 & TIC 142587827 & 0.156 & TIC 143058907 & 0.081 \\
TIC 428257299 & 0.117 & TIC 432128410 & 0.157 & TIC 434097226 & 0.248 \\
TIC 436660156 & 0.210 & TIC 43864479 & 0.176 & TIC 439596199 & 0.108 \\
TIC 441200647 & 0.108 & TIC 450637853 & 0.154 & TIC 452819172 & 0.143 \\
TIC 453097744 & 0.194 & TIC 392536812 & 0.085 & TIC 393943031 & 0.152 \\
TIC 398336271 & 0.122 & TIC 399577123 & 0.147 & TIC 400360028 & 0.070 \\
TIC 401925979 & 0.167 & TIC 401928563 & 0.118 & TIC 402795415 & 0.160 \\
 TIC 406781370 & 0.150 & TIC 411704509 & 0.187 & TIC 415969184 & 0.134 \\
 TIC 336561759 & 0.164 & TIC 339637059 & 0.087 & TIC 348897766 & 0.219 \\
 TIC 354115226 & 0.229 & TIC 356192212 & 0.113 & TIC 365034219 & 0.212 \\
 TIC 36729364 & 0.234 & TIC 377296926 & 0.158 & TIC 380670111 & 0.108 \\
 TIC 38707340 & 0.137 & TIC 387513120 & 0.167 & TIC 387586260 & 0.225 \\
 TIC 198037139 & 0.135 & TIC 257625332 & 0.217 & TIC 289166536 & 0.242 \\
 TIC 332910986 & 0.181 & TIC 426146826 & 0.090 & TIC 293565743 & 0.100 \\
 TIC 29777922 & 0.095 & TIC 314459000 & 0.186 & TIC 315937374 & 0.136 \\
 TIC 316250867 & 0.192 & TIC 316333039 & 0.116 & TIC 316769246 & 0.143 \\
 TIC 320506144 & 0.097 & TIC 330453808 & 0.155 & TIC 258776281 & 0.154 \\
 TIC 260411041 & 0.121 & TIC 261089147 & 0.232 & TIC 261105201 & 0.128 \\
 TIC 261981997 & 0.105 & TIC 268686130 & 0.175 & TIC 279156705 & 0.155 \\
 TIC 279224984 & 0.159 & TIC 199612934 & 0.091 & TIC 204325855 & 0.245 \\
 TIC 211412634 & 0.175 & TIC 233233639 & 0.107 & TIC 237130278 & 0.145 \\
 TIC 250114673 & 0.093 & TIC 255682622 & 0.082 & TIC 257525627 & 0.197 \\
 TIC 156379866 & 0.143 & TIC 159102550 & 0.220 & TIC 164720673 & 0.121 \\
 TIC 170676440 & 0.229 & TIC 193823999 & 0.223 & TIC 47751440 & 0.112 \\
 TIC 49214115 & 0.164 & TIC 53603189 & 0.142 & TIC 55007847 & 0.086 \\
 TIC 56001210 & 0.227 & TIC 5674169 & 0.199 & TIC 63307143 & 0.154 \\
 TIC 63597006 & 0.142 & TIC 66473439 & 0.091 & TIC 671564 & 0.160 \\
 TIC 69838619 & 0.106 & TIC 77178193 & 0.075 & TIC 87751538 & 0.106 \\
 TIC 89428764 & 0.136 & TIC 8989778 & 0.178 & TIC 93053299 & 0.126 \\            
\hline
\end{tabular}	
\end{table*}	

\clearpage
\subsection{Estimation of the third light}

\citet{Hamb+et+al+2024} got an indication of possible third light present in a system. Approximately two-thirds of close binary systems are very likely to be members of triple or multiple star systems \citep{Pribulla+et+al+2006}. Given this, the effects of the third light cannot be neglected. Therefore, we assume that each target is affected by third light. We consider solving for the mass ratio ($q_{l3}$) parameter with the inclusion of third light effects and compare it to the mass ratio ($q_{nol3}$) parameter obtained without considering these effects. In the Phoebe program, the third light is defined as $l_3=F_3/(F_1+F_2+F_3)$. The range of the third light was set from 0 to 1, while all other parameters were maintained as described in Section 3.1. Additionally, a total of 200,000 samples were generated. Similarly, we trained a neural network model (NN$_{l3}$) to establish the mapping relationship between the parameters and the light curves. We assessed the accuracy of light curves generated by the neural network model (NN$_{l3}$) by computing the standard deviation of residuals between the model-generated light curves and the synthetic light curves produced by Phoebe across a sample of 20,000 light curves. The left panel of Figure 11 illustrates that the standard deviations of the residuals are predominantly distributed around $0.0003^{+0.0003}_{-0.0001}$. The relative parameters, including the third light ($l_3$), for 143 contact binaries are determined by employing a neural network (NN$_{l3}$) model in conjunction with the Markov Chain Monte Carlo (MCMC) algorithm. Table 6 presents the parameters of the targets outlined in Table 2. All 143 targets containing the parameter $l_3$, as seen in Table 6, can be found in the GitHub repository \textsuperscript{\ref{fn:github}}. The mass ratios ($q_{nol3}$ and $q_{l3}$) for 143 contact binaries, determined using neural network models (NN$_{nol3}$ and NN$_{l3}$) combined with MCMC algorithms, are compared in the right panel of Figure 11. We calculate the residuals between $q_{l3}$ and $q_{nol3}$. For these residuals, the first quartile $Q_1$ (the 25th percentile) is 0.012, the median $Q_2$ (the 50th percentile) is 0.026, and the third quartile $Q_3$ (the 75th percentile) is 0.05.

It should be noted that this paper makes the following assumption. All the targets in this study are from total eclipse systems, and the majority of them exhibit low mass ratios. Theoretically, the effective temperature of the primary star measured at the instant when the primary star eclipses the secondary star is the most accurate. Nevertheless, owing to the limitations of the survey data, we are incapable of precisely measuring the effective temperature at this particular moment. Our assumption is to regard the effective temperature of the system as that of the primary star. This is because, for targets with low mass ratios, the luminosity of the system is predominantly contributed by the primary star.

\begin{figure}[!ht]
\begin{center}

\begin{minipage}{8cm}
	\includegraphics[width=8cm]{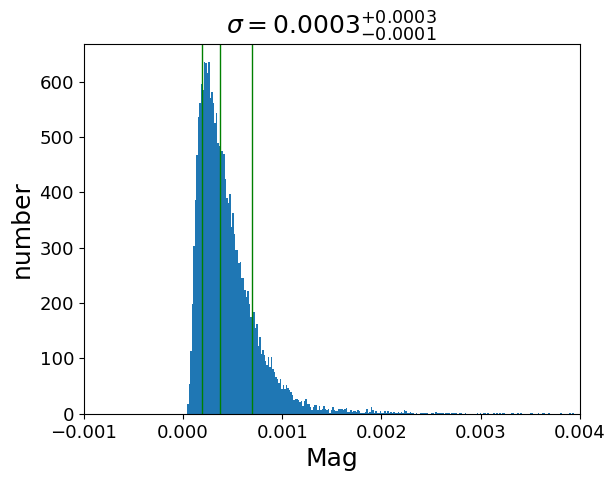}
\end{minipage}
\begin{minipage}{8cm}
	\includegraphics[width=8cm]{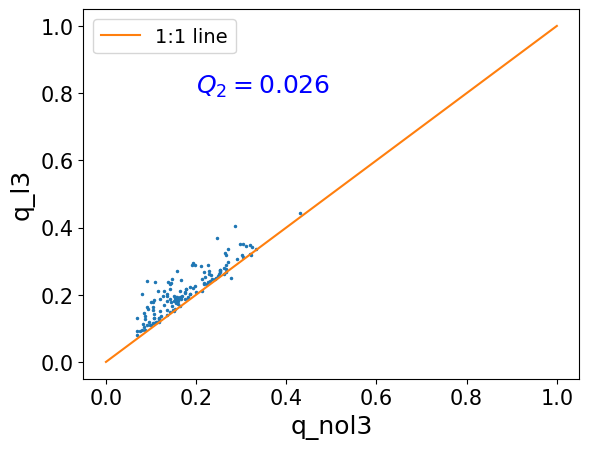}
\end{minipage}

\caption{Left: The distribution of the standard deviation of the residuals between the predicted light curves produced by the model (NN$_{l3}$) and the synthetic light curves produced by Phoebe is depicted. Right: The mass ratios ($q_{nol3}$ and $q_{l3}$) for 143 contact binaries, determined using neural network models (NN$_{nol3}$ and NN$_{l3}$) combined with MCMC algorithms, are compared in the right panel. }\label{Fig21}
\end{center}
\end{figure}

\clearpage
{
\tiny
\begin{center}
\begin{longtable}{ccccccccccccccc}
\caption{The essential parameters, including the third light ($l_3$), of the four contact binary systems from the catalog of 143 targets.}\label{Table 61}\\
\hline\hline                          
name     &$T_1$(K)   &$incl(\circ)$  &$\sigma_{incl}$  &$q$  &$\sigma_{q}$  &$f$  &$\sigma_{f}$   &$\frac{T_2}{T_1}$  &$\sigma_{\frac{T_2}{T_1}}$   &$l_3$  &$\sigma_{l_3}$ &$R^2$   \\
\hline
\endhead
\hline

TIC 671564 & 5607 & 83.350 & 1.635 & 0.172 & 0.010 & 0.744 & 0.039 & 1.064 & 0.004 & 0.058 & 0.036 & 0.995 \\
TIC 89428764 & 6541 & 83.130 & 3.147 & 0.202 & 0.022 & 0.527 & 0.062 & 1.000 & 0.002 & 0.255 & 0.058 & 0.999 \\
TIC 93053299 & 5954 & 85.077 & 3.058 & 0.196 & 0.016 & 0.765 & 0.069 & 1.026 & 0.005 & 0.277 & 0.043 & 0.997 \\
TIC 97328396 & 5935 & 88.402 & 1.132 & 0.307 & 0.004 & 0.519 & 0.018 & 1.015 & 0.003 & 0.027 & 0.008 & 0.993 \\

\hline
\end{longtable}
\end{center}
}

\section{Conclusion}
Totally eclipsing contact binaries can be accurately characterized through photometric methods alone, allowing for reliable determination of mass ratios without the need for spectroscopic data. We have identified 143 totally eclipsing contact binaries from TESS, whose light curves exhibit flat-bottom minima. The trained neural network (NN$_{nol3}$) model accurately generates light curves based on input parameters with an error margin of less than 0.001 magnitudes. Using a neural network (NN$_{nol3}$) model and the MCMC algorithm, we rapidly obtained the fundamental parameters of these contact binaries. By utilizing the relationship between period and semi-major axis using RANSAC algorithm, we estimated  the absolute parameters. Low mass ratio systems (LMR) constitute a particularly intriguing category of contact eclipsing binaries that pose significant challenges to current theoretical models of stability. Additionally, these systems are hypothesized to serve as potential progenitors for the rare optical transients known as red novae. Among our catalog, 96 targets have mass ratios below 0.25, not included in previous studies, suggesting that these targets may be newly discovered low mass ratio system candidates. For these targets, we plan to investigate their dynamical instability in future studies and explore their potential for merging. During the parameter estimation of 143 binary systems, we assume the influence of a third light source. A neural network model (NN$_{l3}$) incorporating this effect is developed. After training NN$_{l3}$, we calculate the residuals between two mass ratio estimates: $q_{l3}$ (considering the third light) and $q_{nol3}$ (neglecting it). Statistical analysis of these residuals using percentiles shows that the 25th percentile ($Q_1$) is 0.012, the median ($Q_2$) is 0.026, and the 75th percentile ($Q_3$) is 0.05.

In addition, there are still some aspects that demand further in-depth research. In Figure 1, we performed a comprehensive analysis of the O'Connell effect exhibited by the target TIC 103656297. A notable finding was that this effect demonstrates a temporal variation. Looking ahead, we are planning to write a research paper that details the time-dependent variations of the O'Connell effect across 143 targets. When it comes to the data in Figure 9, relying solely on photometric solutions to determine the orbital inclination and fill-out may lead to relatively substantial errors. Moreover, the data presented in Figure 11 regarding the influence of the third light on the mass ratio shows an overall high degree of scatter, which is a cause for concern. To tackle these challenges, in our future research, we intend to integrate spectral radial velocity data, aiming to obtain more accurate and reliable results and gain a more profound understanding of the underlying physical mechanisms.

\begin{acknowledgments}
We are very grateful for the data released by the TESS survey (https://archive.stsci.edu/missions-and-data/tess/). This work is supported by the National Key R\&D Program of China (Nos. 2022YFF0711500 and 2023YFA1608300), the Strategic Priority Research Programme of the Chinese Academy of Sciences (Grant No. XDB 41000000), the National Natural Science Foundation of China (Nos. 12103088 and 12433009),  Yunnan Key Laboratory of Solar Physics and Space Science under No. 202205AG070009, Yunnan Provincial Key Laboratory of Forensic Science (No. YJXK005), Yunnan Basic Research Program (Nos. 202201AU070116 and 202501AT070027), and Yunnan Key Laboratory of Service Computing, Kunming, China. We acknowledge the science research grant from the China Manned Space Project under No. CMS-CSST-2021-A10, No. CMS-CSST-2021- A12, and No. CMS-CSST-2021-B10.

\end{acknowledgments}



\end{document}